# A carbon stabilized austenitic steel with lower hydrogen embrittlement susceptibility

Heena Khanchandani[1,*], Stefan Zeiler[1,2], Lucas Strobel[1], Mathias Göken[1], Peter Felfer[1,*]

[1]*Institute for General Materials Properties, Department of Materials Science, Friedrich-Alexander University Erlangen-Nürnberg (FAU), Erlangen 91058, Germany*

[2]*Department of Materials Science, Montanuniversität Leoben, Roseggerstr. 12, 8700 Leoben, Austria*

*Co-corresponding authors: heena.khanchandani@fau.de and peter.felfer@fau.de

**Abstract**
High strength steels are susceptible to H-induced failure, which is typically caused by the presence of diffusible H in the microstructure. The diffusivity of H in austenitic steels with fcc crystal structure is slow. The austenitic steels are hence preferred for applications in the hydrogen-containing atmospheres. However, the fcc structure of austenitic steels is often stabilized by the addition of Ni, Mn or N, which are relatively expensive alloying elements to use. Austenite can kinetically also be stabilized by using C. Here, we present an approach applied to a commercial cold work tool steel, where we use C to fully stabilize the fcc phase. This results in a microstructure consisting of only austenite and an $M_7C_3$ carbide. An exposure to H by cathodic hydrogen charging exhibited no significant influence on the strength and ductility of the C stabilized austenitic steel. While this material is only a prototype based on an existing alloy of different purpose, it shows the potential for low-cost H-resistant steels based on C stabilized austenite.

**Introduction**
The hydrogen embrittlement (HE) susceptibility of high-strength steels [1,2] often leads to a significant loss in their toughness [3]. HE is essentially caused by the diffusion of H through the microstructure which leads to brittle catastrophic failures [4]. The diffusivity of H in austenite is much slower than in ferrite [5,6], thereby lowering the diffusible H content in the austenitic matrix. Austenitic steels hence have been recognized as potential candidate materials for developing the structures resistant to HE [7]. HE is being investigated by the scientific community for over a century [4], however, recently renewed interest is generated by the need of steels for mobile and stationary H storage and a H infrastructure. Stable Ni austenitic stainless steels such as 316 type or 201LN stainless steel [8–12] are used in many H energy applications. However, due to their composition and processing route, they are significantly (ca. 3 - 8 times) more expensive than Fe-C mild steel. It is hence essential to develop a cost-effective C based approach that could be used to generate H resistant austenitic steel.

Ni is often added to partially or fully stabilize the austenitic microstructure in steels such as 304 and 316 stainless steels [13,14], however, Ni is an expensive metal. In 'leaner' concepts such as 201 type austenitic steels [15,16], Ni is partially replaced by Mn and/or N. Yet, the cost is much higher than plain C steels. C is also an austenite stabilizer [14], and can stabilize the austenitic microstructure at room temperature. This typically happens above C contents of 0.6 wt% in plain C steels [17], with higher C contents leading to higher retained austenite fractions [18–20]. This has been an undesired effect in the hardening of tool steels and to the knowledge of the authors, no attempts have been made to systematically use this austenite stabilization mechanism to achieve a fully austenitic microstructure. For the use in non-corrosive H environments such as high pressure H, such a C austenitic steel may however be beneficial, as previous reports suggest that C protects Fe and its alloys from HE [21–24].

In the present study, we demonstrate that C can also be used to push the martensite start temperature to below room temperature, i.e. a full stabilization of the austenite phase at room temperature. This is based on a novel heat treatment of a cold work tool steel as a model system with high C content. We used the cold work tool steel X210CrW12, as it has a high C content of ~2 wt%. However, the high Cr content leads to the formation of Cr carbides, which are undesired in structural materials, as the presence of Cr carbides in the microstructure reduces toughness and ductility.

The overall C content in our model material is close to the maximum soluble C content in austenite of ~2.06 wt% at 1147°C [14]. The steel was austenitized at 1200°C to achieve a high amount of dissolved C and quenched to obtain a fully austenitic structure with Cr carbides in it. We characterized the phase of our microstructure by electron-backscatter diffraction (EBSD) and used atom probe tomography (APT) to quantify the C content in the austenite. The influence of H on the mechanical properties of the material was examined by performing tensile tests on the uncharged specimens and the specimens following the cathodic H charging.

**Materials and methods**

A commercial X210CrW12 tool steel (*MARKS GmbH, Germany*) was selected for the current study, whose bulk chemical composition (wt%) determined by optical emission spectroscopy (OES) is listed in Table 1.

*Table 1. Bulk chemical composition of the commercial X210CrW12 tool steel*

| Element | C | Si | Mn | P | S | Cr | W | Mo | Ni | Fe |
|---|---|---|---|---|---|---|---|---|---|---|
| wt% | 2 | 0.02 | 0.35 | 0.028 | 0.01 | 10.8 | 0.62 | 0.12 | 0.21 | balance |

The soft-annealed tool steel was received in the form of 315.4×1030×2.4 mm$^3$ sheet metal. It was solution heat-treated at 1200°C for 30 minutes in an Ar atmosphere at a heating rate of 10°C/min, which was followed by water quenching to room temperature in order to achieve a stable austenitic microstructure. The heat treatment temperature was chosen based on a Calphad analysis of the phase fractions using Thermocalc. Thermocalc calculations indicated that a significantly higher amount of austenite is formed at 1200°C compared to the usual heat treatment temperature. The steel was cut into tensile specimen geometry of flat coupons with a gauge length of 45 mm using spark erosion. The tensile specimens were mechanically polished with 180-grade, 320-grade, 800-grade and 1200-grade silicon carbide grinding papers. Subsequently, they were mechanically polished with 3µm diamond paste, followed by polishing with colloidal silica suspension.

The tensile specimens were charged with H by cathodic H charging [25–29] using a Gamry Interface 1010B potentiostat. We used a three-electrode setup with a Ag/AgCl reference electrode where the specimen served as cathode at a current density of 40 mA/cm$^2$ and a Pt wire served as an anode. An aqueous solution of 3 wt% NaCl with 0.3 wt% NH$_4$SCN as H recombination inhibitor was used as an electrolyte [30]. The tensile tests were performed on the uncharged and the H-charged specimens at a strain rate of $10^{-4}$ s$^{-1}$ by using an Instron universal testing machine.

The H content in the uncharged and the H-charged specimens was measured by vacuum hot extraction with the H analyzer model LECO TCH 600. The specimens of dimensions 5×5×2 mm$^3$ were charged similarly by cathodic H charging and heated in the H analyzer from room temperature to the melting temperature. The analyses of the released gas following this treatment gives the total H content of the specimen.

A Zeiss Crossbeam 1540 scanning electron microscope (SEM) was used for SEM imaging and EBSD orientation mapping was carried out using an attached Oxford NordlysNano detector. SEM imaging was performed at an accelerating voltage of 20kV and a working distance of 8mm. EBSD was performed at an accelerating voltage of 20kV, a scan step size of 0.5µm, a specimen tilt angle of 70° and a working distance of 16mm. EBSD datasets were analysed by using AZtecCrystal software.

Needle-shaped specimens for APT analysis were prepared by electrochemical polishing using a two-stage process, first, rough polishing was performed in a solution of 25% perchloric acid in glacial acetic acid at 10-20V until the specimen's end was below a micron in thickness. In the second stage, the fine polishing of the specimens was performed in a solution of 5% perchloric acid in butoxyethanol at ~10V. The APT specimens were subsequently cleaned by Ga ions at 2kV in a Zeiss Crossbeam 540 FIB/SEM. APT analysis was performed by using a CAMECA LEAP 4000X HR in voltage pulsing mode at a set-point temperature of 70K, 0.5% detection rate, 15% pulse fraction and 200 kHz pulse repetition rate. Local linear background approximation was used for quantifying the elemental composition by APT using MATLAB R2021a software.

**Results**

Figure 1a reports the initial microstructure of the X210CrW12 tool steel which was received in the soft-annealed condition. It was primarily composed of ferrite (Fe-bcc) with 7.2 vol% of $Cr_7C_3$ and 12.5 vol% of $Cr_{23}C_6$ as shown by its EBSD-phase map in Figure 1a.

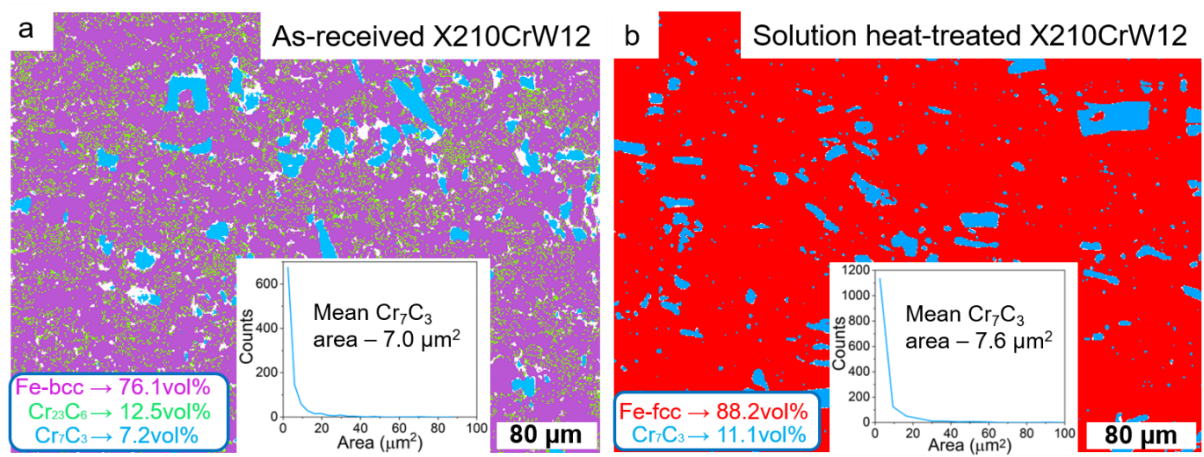

Figure 1. Electron backscatter diffraction (EBSD) phase map of (a) as-received and (b) solution heat-treated X210CrW12 tool steel; inset shows the mean area of $Cr_7C_3$ carbides present in both the materials.

The commercial tool steel was subject to the solution heat-treatment explained above to achieve a stable austenitic structure (Fe-fcc), as shown by its EBSD-phase map in Figure 1b. The vast majority of the microstructure is austenite, no martensite has been produced and ~11 vol% of the microstructure are $Cr_7C_3$ carbide. The average grain size of the material following the heat treatment is approx. 16µm, which is in agreement with many production steels. The blocky appearance of the Cr$_7$C$_3$ carbide suggests that it was retained from the as-received microstructure. The distribution of Cr$_7$C$_3$ carbide size is shown in the insets of Figure 1(a-b) and the average Cr$_7$C$_3$ carbide area in both materials is ~7μm$^2$, which also indicates that the Cr$_7$C$_3$ carbide has been inherited from the as-received material.

APT was used to assess the amount of C that stabilized the austenite. The blue dot in Figure 2a highlights the austenitic matrix region in an SEM image which was analysed by APT, while the corresponding reconstructed 3-D elemental map is shown in Figure 2b. Taking into account all the C species in the data (C$^{1+}$, C$^{2+}$, C$_3^{1+}$, C$_3^{2+}$, C$_4^{2+}$), we quantified that the C content in the analysed specimen is ~1.9 wt% and the Cr content is ~6.4 wt%. This also confirms that the atom probe dataset does not origin from a Cr$_7$C$_3$ carbide. APT analysis hence confirms that approx. 1.9 wt% of C could be dissolved in the austenite matrix by the solution heat treatment. Cr$_7$C$_3$ is a stable carbide [31] where W is not expected to be present. We hence detect a higher W content of 0.9 wt% in the austenite matrix from APT analysis compared to the bulk W content obtained from OES which was 0.62 wt%.

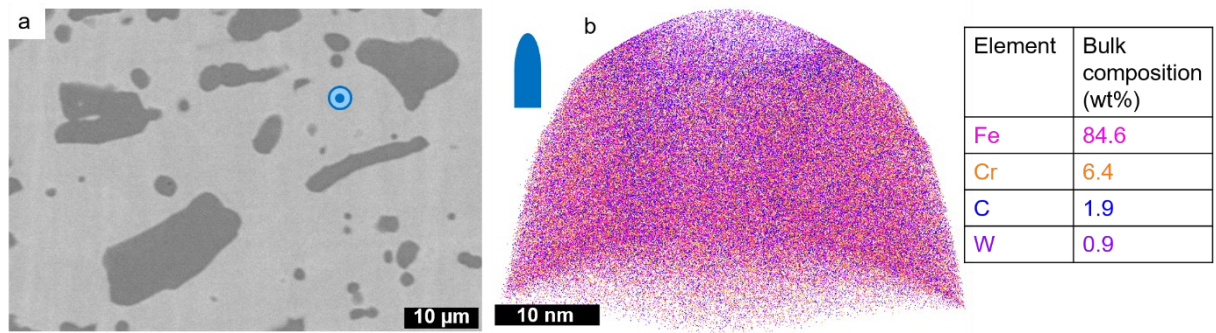

*Figure 2. (a) SEM image highlighting the austenite matrix which was analysed by (b) APT to quantify the C content in the austenite matrix of the solution heat-treated X210CrW12 tool steel.*

Tensile tests were performed on the solution heat-treated specimens with austenite and Cr$_7$C$_3$ carbide in the microstructure with and without H charging. The resulting engineering stress-strain curves obtained from the tensile tests of the uncharged and the H-charged specimens are shown in Figure 3. While the elongation to fracture is very low at ~2% plastic strain, the ultimate tensile strength is remarkably high for a solution heat-treated austenitic steel at ~770 MPa. We attribute the high tensile strength to the solid solution hardening associated to the presence of C in the austenite matrix, which is also suggested by APT results. The stress-strain curves show no significant differences between the as-heat treated and the H-charged material which suggests a lower sensitivity of C stabilized austenitic steel to HE than non-austenitic steels of similar strength levels [32–35]. The H content of the uncharged specimen was 0.2 wt. ppm while the H-charged specimen following the 14h of H charging contained 9 wt. ppm of H. We observe a slight hardening effect in the stress-strain curves of the H-charged specimens which can be ascribed to the solid solution strengthening associated to the presence of H [36,37]. In order to estimate the hardening effect by H, we measured the 0.2% offset yield strength which is defined as the amount of stress that leads to a plastic strain of 0.2%. A straight line parallel to the tensile stress strain curve was drawn at 0.2% strain. The stress level corresponding to the point of intersection of this straight line with the stress strain curve gives the 0.2% offset yield strength as illustrated in Figure 3. The 0.2% offset yield strength is 583 MPa for the uncharged specimen, while it is 613 MPa for the specimen charged with H for 14h. This ~5% increase in the 0.2% offset yield strength is estimated as the contribution of H to the hardening.

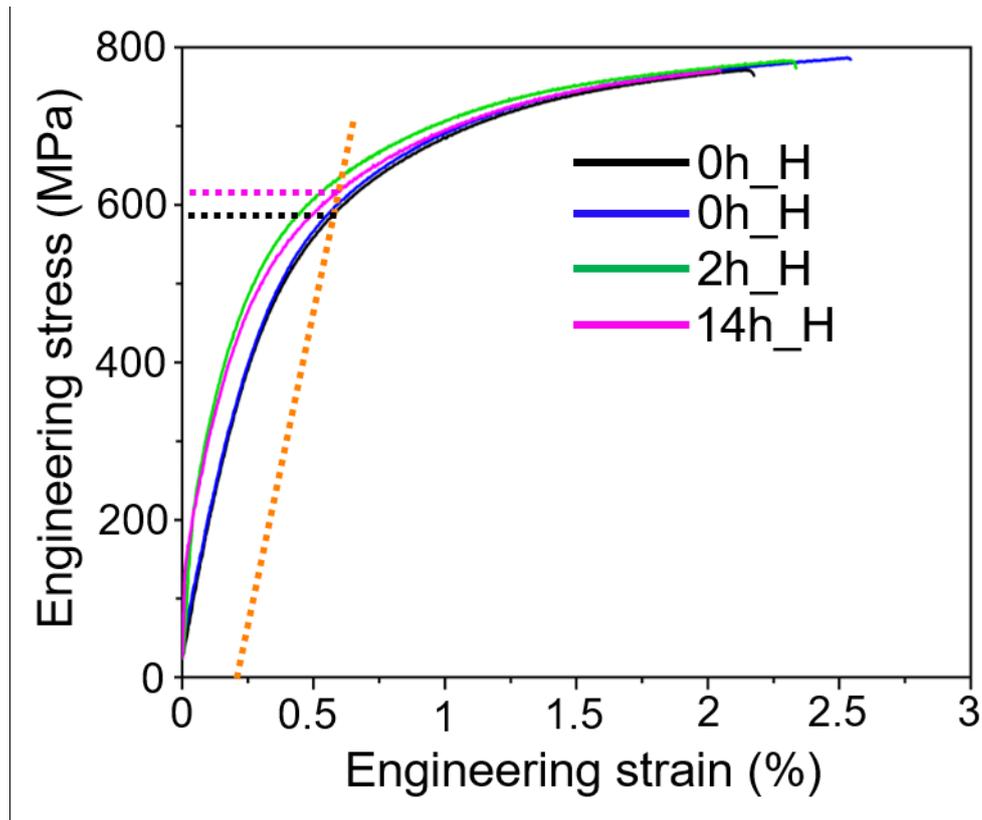

*Figure 3. The engineering stress-strain curves of the uncharged (0h_H) and the H-charged specimens following the 2 hours (2h_H) and 14 hours (14h_H) of H-charging.*

In order to rationalize the origin of the low plastic strains at fracture, we examined the fracture surfaces from the H-charged and the uncharged samples. SEM imaging in Figure 4 evidences the brittle fracture that mostly occurred at the carbide/austenite interfaces with an increased cleavage facet, which is expected from this heterogeneous microstructure. The fracture surface of both the H-charged and the uncharged specimens exhibited similar brittle morphology, which suggests that there are potentially no significant H-induced decohesion processes contributing to the fracture.

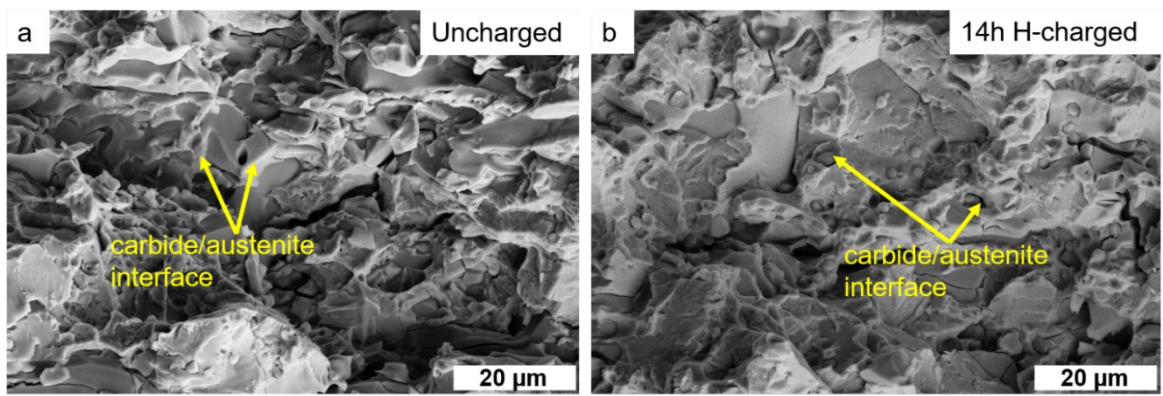

*Figure 4. Fracture surface highlighting the carbide/austenite interface in (a) the uncharged specimen; and (b) the H-charged specimen following the 14 hours of H-charging.*

## Discussion

We were able to create a predominantly austenitic microstructure stabilized with C, with ~11 vol% of $Cr_7C_3$ carbide following the austenitization heat treatment at an unusually high temperature of 1200°C. Although the tool steel used in the current study contains other austenite stabilizing elements such as Ni and Mn, their contents are too low to stabilize an austenitic microstructure in the steel. It demonstrates that the austenitic microstructure in the present work is predominantly stabilized by C. This microstructure shows a high tensile strength with low plastic strain at fracture. The fractography suggests that the low elongation to fracture can be attributed to the presence of carbides in the microstructure. A significant increase in ductility could therefore be expected if a similar material could be produced without the presence of carbides. Unfortunately, we could only obtain the commercial material with a high Cr content at the required high C contents. Therefore, new alloy compositions need to be developed in order to prevent carbide formation and fully explore the ductility potential of this new class of high C austenitic steels.

Beyond the stabilization of the austenite phase, previous reports suggest that C prevents the diffusion of H into the grain boundaries of Fe and its alloys [21,22,24] and hence protects them from H enhanced decohesion. This is one of the most prominent HE mechanisms [38], making the C stabilization route additionally attractive besides the cost and strength benefits. The total elongation to fracture of the material in the current study also did not show any significant change following the cathodic H-charging. Atomistic simulations suggest that the interstitial alloying element C occupies the octahedral interstitial sites in fcc-Fe [39], which are also the favorable sites for H in an fcc-Fe [38,40]. A higher C content in the austenite matrix hence would prevent H from entering those sites which could give rise to a very small H diffusion coefficient, albeit this needs further work for confirmation.

**Conclusion**

We demonstrate that C can be used to fully stabilize an austenitic steel microstructure, which exhibits a high tensile strength of approx. 770 MPa and does not show any significant changes in the ductility following the cathodic H charging. While cold worked Ni or Ni-Mn austenites are also fairly H resistant, the use of C as a stabilizing element has an enormous cost advantage. Since we used a commercially available cold work X210CrW12 tool steel to demonstrate this ability, the resulting microstructure still contained ~11 vol% of $Cr_7C_3$ carbides. APT analysis revealed that ~1.9 wt% of C could dissolve in the austenitic matrix. It was evidenced in the SEM fractography that the carbides led to a reduced strain at failure. We hence cannot yet comment on the ductility potential of high C austenite, however, the current study demonstrates that C based austenitic steels could be promising as a low-cost alternative to stainless steels in H facing applications where no corrosion load is present.

**Acknowledgements**
H.K. and P.F. acknowledge the financial support from the European Research Council under the European Union's Horizon 2020 research and innovation programme (Grant agreement No. 805065). The research of S.Z. has received funding from the European Research Council (ERC) under the Horizon 2020 research and innovation programme (Grant agreement No. 949626). The authors would like to thank Dr. Christopher Zenk for Thermocalc calculations of the phase fractions. We also thank Andreas Kirchmayer and Jan-Oliver Hücking for helping with the measurements of H content in our specimens.